\documentclass[12pt]{article}
\usepackage{epsf}
\begin{document}
\renewcommand{\theequation}{\thesection.\arabic{equation}}
\renewcommand{\section}[1]{\addtocounter{section}{1}
\vspace{5mm} \par \noindent
  {\bf \thesection . #1}\setcounter{subsection}{0}
  \par
   \vspace{2mm} } 
\newcommand{\sectionsub}[1]{\addtocounter{section}{1}
\vspace{5mm} \par \noindent
  {\bf \thesection . #1}\setcounter{subsection}{0}\par}
\renewcommand{\subsection}[1]{\addtocounter{subsection}{1}
\vspace{2.5mm}\par\noindent {\em \thesubsection . #1}\par
 \vspace{0.5mm} }
\renewcommand{\thebibliography}[1]{ {\vspace{5mm}\par \noindent{\bf
References}\par \vspace{2mm}}
\list
 {\arabic{enumi}.}{\settowidth\labelwidth{[#1]}\leftmargin\labelwidth
 \advance\leftmargin\labelsep\addtolength{\topsep}{-4em}
 \usecounter{enumi}}
 \def\newblock{\hskip .11em plus .33em minus .07em}
 \sloppy\clubpenalty4000\widowpenalty4000
 \sfcode`\.=1000\relax \setlength{\itemsep}{-0.4em} }
\newcommand\rf[1]{(\ref{#1})}
\def\nn{\nonumber}
\newcommand{\sect}[1]{\setcounter{equation}{0} \section{#1}}
\renewcommand{\theequation}{\thesection .\arabic{equation}}
\newcommand{\ft}[2]{{\textstyle\frac{#1}{#2}}}

\newcommand{\vgl}[1]{eq. (\ref{#1})}

\thispagestyle{empty}
\hfill KUL-TF-99/20\\
\phantom{}\hfill hep-th/9906080
 
 \begin{center}

\vspace{3cm}

{\large\bf New solutions to an old equation}\\

\vspace{1.4cm}

{\sc F. Denef, J. Raeymaekers and W. Troost}\\

\vspace{1.3cm}

{\em Institute for Theoretical Physics} \\
{\em K.U. Leuven, Belgium} \\

\vspace{1.2cm}

\centerline{\bf Abstract}
\vspace{- 4 mm}  \end{center}
\begin{quote}\small
We present a brief historical overview of the classical theory of a radiating point
charge, described by the Lorentz-Dirac equation.
A recent development is the discovery of tunnelling of a charge through a potential barrier,
in a completely classical context. Also, a
concrete example is discussed of the existence of several physically acceptable solutions for a
range of initial data. We end by pointing out some open problems in connection
 with D-brane and monopole physics.  
\end{quote}
\vfill
\rule{55mm}{.3pt}\newline
{\footnotesize Presented at the conference ``Fundamental interactions: from symmetries to black 
holes'' in honor of Fran\c cois Englert, Brussels, March 25-27, 1999.}
\newpage
\baselineskip18pt
\addtocounter{section}{1}
\par \noindent
{\bf \thesection . Introduction}
  \par
   \vspace{2mm} 
\noindent
More than one hundred years ago, Lorentz determined that the force exerted on
a  particle (charge $e$, no spin, but we will call it ``electron'' nevertheless)
by an electromagnetic field is given by%
\footnote{Even though Lorentz didn't, of course,
we use relativistic notation: $x^\nu$ is the space-time position of the particle, and 
the dot denotes a derivative with respect to its  proper time.}
\begin{equation}
F_\mu=e F_{\mu\nu}(x^\alpha)\dot{x}^\nu \label{Lorentzkracht}.
\end{equation}
The electromagnetic field in \vgl{Lorentzkracht} is to be taken {\em at} the 
position of the charge. However, a  charge generates an electromagnetic field
that is {\em singular}. Hence the expression for the force in  \vgl{Lorentzkracht}
can not be taken literally: somehow it should {\em not} include
the field generated by the particle itself, which would render the expression meaningless.
On the other hand, just leaving it out is not satisfactory either, since it 
would miss an important physical effect: a charge may radiate, and this 
radiation acts back on the motion.
The resolution of this question dates from the beginning of this century, 
and resulted in an equation, nowadays called the {\em Lorentz-Dirac equation} (LDE),
that takes into account this back-reaction.
In the next section we review some highlights in the history of this equation.
It can be formulated as a {\em third} order differential equation, which 
entails some peculiar features of the associated particle trajectories.
Among these peculiarities we mention (in section~3
) the well known runaways,
and call 
attention to the fact that there is an indeterminacy in the theory: for given 
initial conditons (combined with physical asymptotic conditions), 
{\em multiple} trajectories may be possible.

There has been a steady (if limited) interest in this subject that never 
completely faded. It may come therefore as a surprise that recently we 
discovered a whole {\em new class of solutions}, which, if not very relevant 
experimentally, nevertheless poses some intriguing theoretical questions. This 
class describes {\em tunnelling}, in a completely classical context: it is possible for a charge
to traverse a classically forbidden region, provided this is accomplished in a sufficiently short
time. We illustrate this in section~4
. 
We end with an assortment of additional remarks connecting this old chapter to 
more modern developments, like monopoles and D-branes.

\setcounter{equation}{0}
\section{The Lorentz-Dirac equation: a short historic overview}
\noindent
Here is the Lorentz-Dirac equation
\footnote{See \cite{general} for general references.}:
\begin{eqnarray}
\stackrel{..}{x}^\mu&=&\quad\frac{e}{m\,c}F^{\mu\nu}\stackrel{.}{x}_\nu
  \ \ +\ \ \frac{2 e^2}{3 m \,c^3} (\frac{\stackrel{..}{x}^2\stackrel{.}{x}^\mu}{c^2}
         +\stackrel{...}{x}^\mu)
\label{LDE} \\[2mm]
&&\mbox{ \hspace{.9cm}      (1) \hspace{2.6cm}   (2) \hspace{.7cm}    (3)   \nonumber         
}%
\end{eqnarray}
The first term is the Lorentz force. The second term results from 
the non-relativistic theory of the radiation reaction as developed by Lorentz 
and Abraham in the (eighteen-)nineties (the electron was discovered experimentally at the 
end of that period). It takes into account
the energy loss due to radiation, as should be familiar.
It was also found to be in agreement with considerations of extended models
 for a charged particle (Poincar\'e), in the limit that its size tends to 
 zero. This involves a classical renormalisation of the mass of the charged particle.
 That the third term in \vgl{LDE} is a necessary is clear when one checks that $\dot{x}^2=1$
 is preserved. It follows from a proper relativistic treatment as given first by Schott (1915),
  and is (perhaps surprisingly) called the  ``Schott~term''. 
 
Facing some pecularities of the solutions of \vgl{LDE} --- to which we come back soon---,
over the years, several alternatives have been (re-)proposed. 
For point electrons, these typically run into difficulties,
with energy-momentum conservation. Indeed, Dirac showed\cite{Dirac} in 1938 that 
the equation as it stands follows from very general considerations on the 
conservation of the energy-stress tensor everywhere up to the immediate 
vicinity af the electron world line.%
\footnote{Dirac's derivation eventually attached his name to the equation.}
Therefore, the only alternative is an extended model, but this is not without problems.
 A rigid model is incompatible with special relativity. Dirac's own deformable 
 model  of 1962 (which he hoped would ``explain'' the muon) was shown in 1978
 \cite{Hasenfratz} to be instable.%
 \footnote{This is illustrative of the slow but steady progress
  in this outpost of classical physics. }
The situation was  summed up in the sixties in the classical work of 
Rohrlich\cite{Rohrlich}\footnote{At present, it is still an excellent starting point, 
even though it is incomplete (see further). The paper by T. Erber\cite{Erber} 
contains complementary information.}

In the meantime, with the improved understanding of the physical (i.e. quantum mechanical)
electron, the quantum field techniques were applied to this problem, without shocking results:
the equation (\ref{LDE}) also follows if one\cite{Coleman} applies the new renormalisation 
techniques. A singularly missing piece of the  puzzle however is a direct connection 
of the equation to quantum electrodynamics.  

\setcounter{equation}{0}
\section{Solutions\label{oplossing}}
\noindent 
Now we look at some solutions of \vgl{LDE}.
Already for the motion of the electron in a region {\em without} external
electromagnetic fields there is a surprise: apart from the familiar linear solutions
there is a solution with exponential rising rapidity,
like $\dot x^\mu =c\, (\cosh Y, \sinh Y, 0, 0)$ with 
$Y=\exp(\frac{\tau}{\tau_0})$. Here, $\tau_0=\frac{2 e^2}{3 m \,c^3}$ is the 
natural time unit for this ``runaway'' phenomenon, its value for an electron
is $\simeq 0.62\ 10^{-23}\ \rm s$. This phenomenon is not observed in nature. 
In quantum theory, for an electron, a typical timescale would be 
given by the Compton radius divided by the light velocity,
$\frac{\hbar}{m\,c^2}=\tau_0 \frac{3}{2\alpha}$. Therefore, a way out is to
state that one should not take seriously the classical theory down to this 
timescale. Be that as it may, one should eliminate these solutions from the classical theory.
This is achieved by an asymptotic (large time) condition: one decrees that,
if the particle ends up in a force free region, it does not accelerate asymptotically.
While eliminating the 
runaways, this procedure in turn gives rise to two new surprises.
The first appears if we solve \vgl{LDE} for a particle moving into the vicinity
 of a potential step: it starts to accelerate
{\em before} it reaches the edge of the step.
 The pre-acceleration time, i.e. the characteristic time of this onset of the acceleration,
 is again $\tau_0$. Although some may consider this a 
 violation of causality, the phenomenon is less shocking if one takes into 
 consideration that the Coulomb field, with its singularity at the position of 
 the particle, is noticed in the region of the force field long before the particle arrives there.
The second is the existence of {\em multiple} solutions of \vgl{LDE} for given 
initial position and velocity. Although in \cite{Rohrlich} the question of uniqueness was
called ``one of the most important unsolved problems of the theory''
\cite{Rohrlichcit}, the older literature already contains examples of 
non-uniqueness, as well as conditions for uniqueness\footnote{ which time does not 
permit us to go into, but see \cite{Haag}.}. During the seventies (see for example
\cite{Huschilt,Carati}) this aspect 
was stressed at the occasion of (often numerical) studies of the LDE in 
various circumstances. A recent addition\cite{SSC} is the proof of uniqueness in a constant magnetic field,
a study issued from the ill-fated SSC-laboratory. More on this non-uniqueness topic
 in the next section.

During the seventies, an interesting interpretation of \vgl{LDE} was worked out (see 
\cite{Teitelboim}) in terms of a {\em bound momentum} 
\begin{equation}
p^\mu= m \dot x^\mu - \frac{2 e^2}{3 \,c^3} \stackrel{..}{x}^\mu
\end{equation}
which contains, apart from the usual velocity term, also a contribution from the 
acceleration when the particle is charged. This quantity is a property of the particle's
motion at a fixed time, and corresponds to a consistent split of the 
energy-momentum tensor in a part ``bound'' to the electron, and a part 
radiating away. In terms of this momentum, transfering the third term on the right hand side
of \vgl{LDE} to the left,  the equation of motion assumes a completely 
conventional interpretation: the rate of change of the momentum is equal to 
the applied force, where the latter consists of two pieces, the externally 
applied force (term (1)) and the radiation reaction force (term (2)). This interpretation is 
very useful, as we shall see, to guide the intuition.

\setcounter{equation}{0}
\pagebreak[4]
\section{Tunnelling{\label{tunnelsektie}}}
\noindent 
In spite of the venerable history of the subject, a whole new 
class of solutions lay undiscovered until recently\cite{Wij}. The surprising fact is 
that, according to \vgl{LDE}, a classical particle may cross a potential 
barrier provided it can do so in a time of the order of $\tau_0$. The 
demonstration is completely elementary for a particle crossing a rectangular potential 
barrier, in one dimension. The equation, in natural units, reduces to
\begin{equation}
\stackrel{.}{p}=\stackrel{.}{v}-\stackrel{..}{v}=F\label{NRLDE}\,.
\end{equation}
This may be viewed as the non-relativistic approximation (where $v$ is the velocity),
but the equation is in fact exact also relativistically if $v$ is taken to be the rapidity.
\footnote{The presentation we follow in the sequel is non-relativistic, but 
only because of the more transparant relation between distance and velocity in 
that case. None of our conclusions depend on it.}
The electron  only experiences a force when
crossing the boundaries of the regions of constant potential.
The solutions in the separate force-free regions, which are easy to write
down explicitly, are connected using the following matching 
condition%
\footnote{This can be checked using the formal equation
$\stackrel{.}{p}=\Delta V \cdot \delta(x)$. The validity of the rectangular
barrier idealisation is discussed in \cite{Wij}.}
 on the momentum, or the acceleration, the position and  the velocity being
continuous:
\begin{equation}
\Delta p = -\Delta \dot{v} =- \Delta V /v \ .\label{match acc}
\end{equation}
This results in the following set
of equations relating the initial and final velocities to the
time $T$  spent in the barrier region of  width $w$ and height $V$:
\begin{eqnarray}
w&=& v_f T- \frac{V}{v_f}(e^{-T}-1+T) \ ,\nonumber\\
v_i&=& v_f -\frac{V}{v_f}+\frac{V}{v_f-\frac{V}{v_f}(1-e^{-T})}\ .
\end{eqnarray}
Although the analysis of these equations
in general is not very difficult, it becomes
particularly simple when the final electron energy is equal to half the
barrier height, $V=v_f^2$. An explicit example is, for $V=144,w=3$:
\begin{equation}
\begin{array}{rcrcl}
x=&-7 (e^t-1)+16 t\mbox{    }    &   &t&<0     \ ,  \\
  & 9 (e^t -1)                       & 0<&t&<T \ ,  \\
  & 3+12 (t-T)                   & T<&t&       \ ,
\end{array}        \label{spesol}
\end{equation}
with $T=\log 4/3$. The matching condition eq.(\ref{match acc}) implies jumps
of $16$ and $-12$ units in the acceleration at $t=0$ and $t=T$.
Tunneling occurs, the initial energy is equal to 128,
a fraction 1/9 below the barrier height.

The motion can be described intuitively by following the bound momentum, 
\vgl{NRLDE}, while the electron crosses the barrier.  Its value is piecewise
constant, but the velocity component of the momentum and the acceleration component may 
change continuously, $v=p + {\rm constant. }\exp(\tau)$.
 This is akin to the pre-acceleration phenomenon.
In the regions outside the barrier the momentum is equal to its asymptotic value.
Under  the  barrier  itself the value is $v_f-V/v_f$. Thus, the momentum may be in
the same direction {\em or opposite} to the velocity.\footnote{The explicit 
example given above is special in that the momentum vanishes in the intermediate region.}
Within a time of the order of the pre-acceleration time $\tau_0$ this situation
may be ``rectified'', bringing momentum and velocity back in line. However, in the 
meantime the electron may have reached the other side of the barrier. In that 
case, it tunnels through.

Whereas the details of this example are of course special,
the tunnelling phenomenon is actually quite generic. A decisive parameter
is the width of the potential. If the electron can cross the barrier
within a time of order 1, i.e. the pre-acceleration time $\tau_0$, tunnelling
occurs. The  smaller  the  width,  the  wider  the  range  of  initial
velocities  for  which  the  electron  will tunnel.

\begin{figure}
\epsfxsize=\textwidth
\epsffile{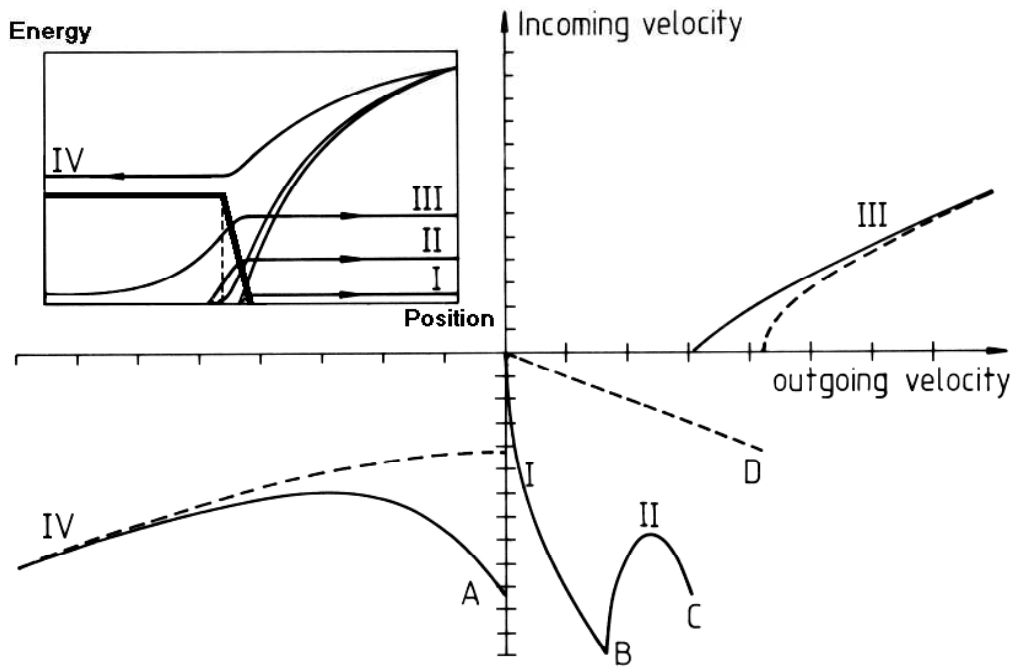}
\caption{Plot of the initial velocity vs. the final velocity
for the solution of the Lorentz-Dirac equation in a linearly rising
step potential (note the difference in scale).
The dotted lines leave out the radiation reaction.
The inset shows the potential and kinetic energy as a function
of position for four representative examples.
The four types of motion are discussed in the text. For the plots,
a step height $V=9$ was used, and a slope width $\epsilon=0.5$.
}
\label{rampfiguur}
\end{figure}

Since a sudden step in the potential actually corresponds to an infinite 
force, one may well be worried that the tunnelling solutions are an artifact of 
that unphysical feature. This is not the case. We have made a detailed 
investigation of what happens at a (steep) ramp, a potential step being
an idealisation of this. The results are shown in figure~\ref{rampfiguur}, taken from~\cite{Wij}.

In keeping with the solution strategy that corresponds to integrating the 
equations backwards (so as to use the asymptotic condition with maximal 
efficiency), the plot is of incoming velocity (rapidity) {\it vs.} outgoing 
velocity. The  inset shows the four types of motion around the ramp,
the main figure shows the relation between the asymptotic velocities.
The branch labeled IV is investigated in very great detail in~\cite{Parrott}.
The branch II, in the limit of an infinitely steep potential, is investigated 
numerically in~\cite{Gull}, where it is concluded that for a certain range of 
initial velocities {\em no} solution exists. Whereas this is true for the 
infinitely steep case, we see that the missing range of small initial 
velocities is covered by our branch I, and it is therefore an unphysical 
feature of the infinite -force approximation. We see clearly that it 
corresponds to the particle turning back under the sloping region, which is 
reduced to zero width in that approximation. For more details on figure~\ref{rampfiguur}, we 
refer to the original paper. The complete figure is a good illustration of the 
fact that for specific initial velocities several different solution may 
exist: up to five in this case, on branches I,II~and~IV. A complementary 
remark is that actually for {\em all} initial values a solution is found. We 
do not know of a proof of this fact in any generality, nor have we attempted 
to construct one. It would clearly be worrying if some initial velocities 
would turn out to be impossible, but happily this is not the case here.

\setcounter{equation}{0}
\section{Remarks}
\noindent 
For the electron, as mentioned in section~3
, the pre-acceleration time
(times $c$) is a factor $\alpha$ smaller than the Compton wavelength, and 
therefore one expects quantum effects to mask any classical effect on that 
timescale. Lacking any solid investigation, it is an open question whether a
full quantum theory, QED for instance, does give rise to the LDE in some 
suitable limit. If so, the classical tunnelling solutions we found
(and the undeterminacy of final velocities) should have a quantum counterpart and
correspond to physical features of the electron theory.

For magnetic monopoles, this argument may be reversed\footnote{This remark 
discusses a point raised by G.~'t Hooft at the meeting.}: since in that case
the size of the coupling is the inverse of the electric coupling, quantum theory
does not bail out the classical theory, so the ``strange'' features of the LDE
and its solutions should find a ``resolution'' within the classical theory.
Happily, purely classical field theories are available where a monopole exists
\cite{monopole} as a soliton. The monopole is extended, and the fields have
a {\em finite} energy so that no infinite renormalisation is needed. Since 
another length scale comes in due to the finite extension, the final form of 
the effective equation of motion does not have to reproduce \vgl{LDE} exactly.
It would be interesting to investigate%
\footnote{
In \cite{BakMin}, in a the nonrelativistic approximation, 
a correction to \vgl{LDE} is found proportional to {\em fourth} order 
derivatives.}
to what extent it does, and especially
what happens to the unexpected features like tunnelling.

As soon as a substructure of the electron is considered, other possibilities 
lie open. Apart from mechanical models, probably the oldest variation on this 
theme is the modification of electrodynamics known as Born-Infeld 
electrodynamics. In currently fashionable developments
in connection with string and M-theory, actions of the Born-Infeld type govern
the dynamics of D-branes. 
Point charge solutions in the classical theory
got a new lease of life\footnote{Old solutions to new equations.}
 following their use\cite{CallanMaldacena,Hashimoto}  for the description
of strings ending on these branes. In such an 
interpretation, an electrostatic self-energy divergence is re-interpreted as 
the (infinite) length of a string stretching from a 3-brane to infinity.
A finite energy solution is obtained by having the string end on another 
brane, at a finite distance. Therefore, from a vantage point on the brane,
this configuration provides a finite 
relativistic model for a point charge as well. A serious drawback, for our 
purposes, of the Born-Infeld dynamics is its non-linearity. It is not clear 
whether a consistent split can be made between the field of the point charge 
and the external fields, or between the energy-momentum ``bound'' to this 
charge and the energy -momentum of radiation: these were instrumental in the 
development of the Lorentz-Dirac equation.
It would certainly be amusing if string theory would not only open new vistas on
traditional field theory, but would improve the status of classical electron theory
as well.

\vspace{.75cm}

\noindent {\bf Acknowledgements}: F.D.  acknowledges the financial support of
the F.W.O. (Belgium). W.T. thanks the organisers of the meeting in honor of F. 
Englert for the occasion to present this work.



\vspace{.75cm}


\end{document}